\documentclass[12pt,preprint]{aastex}
\newdimen\minuswidth    %define @ width of minus sign for tables
\setbox0=\hbox{$-$}
\minuswidth=\wd0
\catcode`@=\active
\def@{\kern\minuswidth}
\newdimen\digitwidth    %define ! a one digit width for tables
\setbox0=\hbox{\rm0}
\digitwidth=\wd0
\catcode`!=\active
\def!{\kern\digitwidth}

\shorttitle{On the Iron content of NGC 1978}   
\shortauthors{Ferraro et al.}

\begin{document} 
  
\title{On the Iron content of NGC 1978 in the LMC:
a metal rich, chemically homogeneous cluster\footnotemark[1]}
\footnotetext[1]{Based on 
observations collected at the Very Large Telescope of the
 European Southern Observatory
(ESO), Cerro Paranal, Chile, under programme 072.D-0342 and 074.D-0369.}

\author{Francesco R. Ferraro\altaffilmark{2},
Alessio Mucciarelli\altaffilmark{2},  
Eugenio Carretta \altaffilmark{3},
Livia Origlia\altaffilmark{3}}
\footnotetext[2]{Dipartimento di Astronomia Universit\`a di
Bologna, via Ranzani 1, I--40127 Bologna, Italy;
francesco.ferraro3@unibo.it;alessio.mucciarelli@studio.unibo.it}   
\footnotetext[3]{INAF--Osservatorio Astronomico di Bologna, via Ranzani 1, I--40127
Bologna, Italy; eugenio.carretta@bo.astro.it, livia.origlia@bo.astro.it}
\medskip

\begin{abstract}  
We present a detailed abundance analysis  of giant stars in NGC 1978,  a
massive, intermediate-age stellar cluster in the Large Magellanic Cloud,
characterized by  a high ellipticity and suspected to have a metallicity
spread. We analyzed 11 giants, all cluster members,  by using high
resolution spectra acquired with the UVES/FLAMES  spectrograph at the
ESO-{\it Very Large Telescope}.  We find an iron content of [Fe/H]=-0.38
dex with very low $\rm \sigma_{[Fe/H]}=0.07$ dex dispersion,  and a mean
heliocentric radial velocity $\rm v_r=293.1\pm0.9$ km/s and a velocity 
dispersion $\rm \sigma_{v_r}=3.1$ km/s, thus excluding the presence of a
significant  metallicity, as well as velocity, spread within the
cluster.  
\end{abstract}  

\keywords{Magellanic Clouds --- globular clusters: individual (NGC~1978) ---
techniques: spectroscopic --- stars:abundances}   

\section{Introduction}   
\label{intro} 
 
The Large Magellanic Cloud (LMC) is the nearest galaxy of the Local Group with 
a very populous system of Globular
Clusters (GCs) that cover a wide range of metallicity and age. 
At least three main populations can be distinguished, namely an
old population, coeval with the Galactic GC system, an
intermediate population (1-3 Gyr) and a young one ($< 1 $Gyr). 

Despite its importance, there is still a lack of systematic and homogeneous works aimed 
at determining the accurate chemical abundances and abundance patterns of the LMC
GC system. 
Starting from the first compilation of   metallicity by \citet{sp89},
the most systematic analysis remains the work by \citet{ols91}, who 
estimate the metallicity of 70 LMC clusters using the Ca II triplet. 
Other metallicity determinations 
are based on the Lick spectral indices \citep{fre98}, 
integrated infrared (IR) spectroscopy \citep{oo98} or derived 
from Str\"omgren 
\citep{dirsch00, larsen00}
and Washington \citep{bic98} photometry.\\ 
Detailed chemical abundances of LMC GCs from medium-high resolution spectroscopy are still scarce.
\citet{hill00} (hereafter H00) measured Fe, O, Al, Ca and Ti abundances of
a few giants in four GCs (namely NGC 1866, NGC 1978, ESO 121 and NGC 2257),  
by using high resolution UVES spectra. 
\citet{kor00} and \citet{kor02} measured a few B stars in 4 young LMC clusters and 
inferred chemical abundances of 
Fe, C, N, O and other $\alpha$-elements \citep[see also][]{ric89}.
\citet{smi02} measured 4 giants in NGC~1898 and NGC~2203 
and obtained accurate abundances of Fe, C, N, O, Na, Sc, Ti.
Results about the chemical 
composition of 4 old LMC GCs (namely 
NGC 1989, NGC 2005, NGC 2019 and Hodge 11) are presented by 
\citet{j04},
based on high-resolution spectra taken with MIKE at the Magellan telescope.\\

In this letter we present the first results of an undergoing project aimed at screening the 
chemical composition of a complete sample of LMC GCs and their surrounding field 
populations, by using UVES/FLAMES.
The major goal of our work is to derive a new homogeneous metallicity scale 
based on high resolution spectroscopy together with 
a detailed description of the abundance patterns of  key metals as $\alpha$, 
iron-group and neutron-capture elements. \\
The first target observed in our survey is NGC 1978.
This intermediate-age \citep[$\approx$3.5 Gyr,][]{girardi} 
cluster is very massive   
\citep[$\sim2\cdot10^{5} M_{\odot}$,][]{wes97}
and located in a high density stellar region,
about 3.5$^{\circ}$ north of the bar field.
It also shows a peculiar, 
very high ellipticity   \citep[$\epsilon=0.3$,][]{gh80, fisher92}. 
The multicolor BVRI photometry by \citet{alcaino} has shown
a broad Red Giant Branch (RGB), consistent with a metallicity spread 
[Fe/H]$\sim$0.2 dex. On the basis of this evidence, the authors suggested
the possible existence of 
two different sub-populations as the result of a merging.
This scenario was furtherly supported by 
H00 who analyzed the high resolution spectra of two 
giant stars located in the south-east region of the cluster. 
They found [Fe/H]=--1.1 and -0.82 dex, 
with a significant star-to-star difference ($\Delta$[Fe/H]$\approx$0.3 dex).
However, the same stars were previously observed by \citet{ols91}, 
who found [Fe/H]=-0.46 and -0.38, i.e. a much higher (by a factor of $\approx$3) 
metallicity and a much smaller ($\Delta$[Fe/H]$\approx$0.08 dex) star-to-star difference.

In order to better understand the formation and evolution of NGC 1978,
a detailed high resolution spectroscopic study of a significant sample of 
cluster stars is  needed.
Here we present the detailed abundance of Iron 
for 11 giants in NGC 1978. 
  
\section{Observations \& Spectral Analysis
\label{obs}}

\begin{figure}[t!]
\epsscale{1.0}   
\plotone{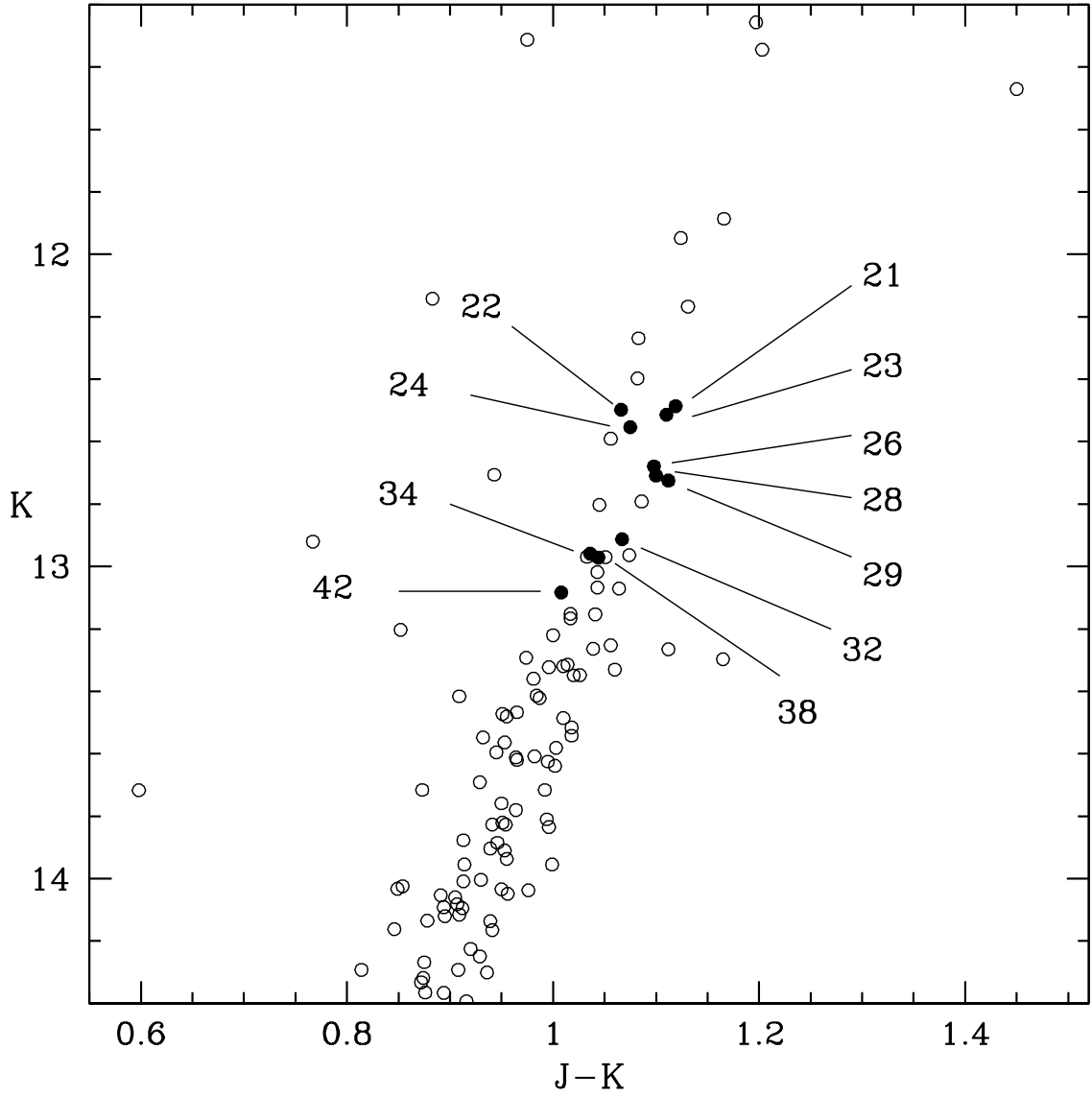}
\caption{IR (K,J-K) color-magnitude diagram
of NGC 1978 from \citet{m06}. 
The 11 program stars (black points) are labeled accordingly to their identification 
number in Table~1.
}
\label{cmd}
\end{figure}

In order to establish whether a metallicity spread is present throughout NGC 1978,
11 RGB stars 
were observed in two different runs on October 2003
(ESO Program 072.D-0342(A)) and February 2005
(as a back-up programme within the ESO Program 074.D-0369(A)).
We used the multi-object spectrograph UVES/FLAMES \citep{pasquini02}, mounted 
at the Kueyen 8 m-telescope (UT2) of the ESO Very Large Telescope (VLT).
The UVES set-up (RED ARM, centered at 5800 $\mathring{A}$) provides a wavelength 
coverage of 4800-6800 $\mathring{A}$ and a resolution R$\sim40000$.
The spectra have been acquired in series of 4-6 exposures of $\approx$45min each, 
flat-field corrected  and 
average-combined together for a total exposure time of 3-5 hrs. 
The final spectra have typical  $\rm S/N\ge$40. 
The selection of the target stars is based on our high quality near-IR  
photometry of the cluster by using SOFI mounted at the ESO-NTT
 \citep{fer04a,m06}.
Fig.~\ref{cmd} shows the position of the 11 giants in the IR  K,(J-K) Color Magnitude
Diagram (CMD).
The stars are also well distributed within the cluster area, as shown in Fig.~\ref{map}.

Fig.~\ref{spec} shows an example  
of the final spectra used for the spectral analysis.  
From the measured radial velocity (see Table 1) we find that
all the 11 stars are cluster members, with    
a mean heliocentric velocity $<v_{r}>$=+293.1$\pm$0.9 km/s, 
and a velocity dispersion $\sigma$=3.1 km/s,
in excellent agreement with the 
$v_{r}=+$292.4$\pm1.4$ km/s,
previously determined by 
\citet{ols91}.
 
The analysis of the chemical abundances was performed 
using the ROSA package \citep{g88}. 
 The line equivalent widths (EWs) of the observed spectra have been measured by 
Gaussian fitting the line profiles, adopting a relationship between EW and FWHM 
\citep[see e.g.][]{brag}; an iterative clipping average over 
a fraction of the highest spectral points around each line has been applied to derive a local continuum.
   The details of the line list 
and the corresponding atomic parameters are given in \citet{g03}.
The stellar temperatures ($T_{eff}$) have been estimated using the IR 
(J-K) color and the transformations by 
\citet{alonso99,alonso01} and \citet{m98}. 
Since the difference between the two temperature scales
 in the cool regime is always $<50$ K, we adopted the average of the two values.
Gravity has been estimated accordingly to the location  of the stars
in the CMD and using 
a theoretical isochrone of $\sim$3 Gyr and Z=0.008 from \citet{cariulo04},
by assuming a stellar mass of 1.37 $M_{\odot}$, 
a distance modulus of $(m-M)_0$=18.5 \citep{vdb98}, a reddening of E(B-V)=0.1 \citep{perss83} 
and the interstellar extinction law defined by \citet{rl85}. 
For the bolometric corrections we 
used those computed by \citet{m98}.  
Note that a slightly different choice of the isochrone metallicity has a negligible impact 
on the inferred stellar gravity: indeed, by varying the former  by a factor of 2, 
the mass  changes by $\approx$0.03 $M_{\odot}$ which translates into a 
gravity variation 
of $\approx$0.01 dex. 
Conversely different assumption for the cluster age 
can have some impact, we find that a 1 Gyr age variation implies
a  0.05 dex gravity variation.
Accordingly to \citet{mag84} prescriptions, 
the microturbulence velocity $v_t$ (see Table 1) 
is obtained  by removing the residual trend of the derived 
FeI abundances with the predicted line strengths  X (defined as $\log{gf}-\theta\chi$), 
using a large number (typically 70-80) of FeI lines for each star.
ATLAS model atmospheres with convective overshooting by \citet{kur} are used to 
perform the abundance analysis.

Table 1 shows the adopted atmospheric parameters and 
the values of [Fe/H]I
\footnote{We adopt the usual spectroscopic notation: [A]=log$(A)_{star}$-log$(A)_{\odot}$ for any 
abundance quantity A; log(A) is the abundance by number of the element A in the standard scale 
where log(H)=12.} and [Fe/H]II for all the program stars. The $N_{FeI}$ and $N_{FeII}$
number of lines 
used to derive the abundance
are also listed.
We adopt reference solar log n(FeI)=7.54  and n(FeII)=7.49 for neutral and ionized Fe, 
respectively \citep[see ][]{g03}. Given the low temperature of the observed stars and in order to
avoid spurious effects  due to line blending, only a few safe lines   were used to derive the FeII
abundance. In particular, for three stars (namely, NGC1978-21, NGC1978-34, NGC1978-23) no good
lines  are available.\\ 
Plots reported in Fig. 4 represent a test to the validity of our analysis.
In particular, the absence of any trend of $\Delta(FeI)/\Delta(\chi)$  (where $\chi$ is  the
excitation potential) with respect to $T_{eff}$ (mid panel of Figure 4)  supports the reliability
of our temperature scale\footnote{We estimate that the typical $T_{eff}$ derived from excitation
equilibrium  should be lower by $\delta T\sim75\pm100$ K with respect to the photometric  
estimates. This systematic turns out to be comparable with  internal errors in $T_{eff}$.}.
Similarly, the absence of trend in upper panel is a good proof of the  correctness of our
microturbulent velocities. We underline this point because H00 estimated (for similar stars in this
cluster) a larger value (tipically $v_t\sim1.9$ km/s). This difference is clearly due to the
different metodology used to  calculate this parameter: H00 used the observed EWs and not, as we
do, the expected  line strengths.

\section{Results and discussion} 

Our spectroscopic analysis based on 11 cluster member stars 
provides an average iron abundance from neutral FeI lines of 
[Fe/H]I=--0.38$\pm$0.02 dex
and [Fe/H]II=--0.26$\pm$0.02 dex from singly ionized lines. 
The overall metallicity dispersion is $\sigma$=0.07 dex.
The overall error budget in  [Fe/H] has been
computed accordingly to the uncertanties in the 
adopted atmospheric parameters and in the
measured EWs. 
Uncertainties in temperatures (typically $\pm$ 60 K)
are estimated by taking in account the
errors of the infrared colors (typically $\delta(J-K)\sim 0.02$ mag) and  
reddening ($\delta E(B-V)\sim 0.02$ mag). 
The uncertainty in gravity ($\pm$ 0.08 dex) is obtained by 
quadratically summing  uncertainties in temperature, 
in distance modulus and in bolometric correction.
1$\sigma$ random error ($\pm$ 0.11 km/s) in
microturbolent velocity has been estimated 
from the slope of the 
abundance/line strenght relation.   
The internal errors in [A/H] are typically less than 0.10 dex.
Finally, the contribution of the EW 
measurement uncertainties
to the abundance error budget  was estimated by dividing the average rms 
scatter of FeI lines (assumed to represent the error of each 
individual line) by the square root of the  number of lines.
Considering all these errors sources we obtain a total uncertanty of $\pm$0.07 dex for
[Fe/H]I and $\pm$0.17 dex for [Fe/H]II,  fully consistent with the (low) cluster metallicity
dispersion.
This confirms the high homogeneity level in iron content 
of this cluster\footnote{A further test about the validity of our analysis was performed. 
We divided our sample in two sub-groups: the first included stars with $T_{eff}\sim3750 K$ 
and the second with  $T_{eff}\sim3850 K$; only the coolest star (NGC 1978-23) is excluded. 
The spectra of stars in each group have been summed and high S/N combined spectra 
were obtained. We repeat the abundance analysis described above, using the 
average atmospheric parameters for each group. The resulting [Fe/H] from these combined spectra 
is in excellent agreement with the the iron content derived from individual stellar spectra,  
the difference being $\le0.03$ dex.}. 

Our average metallicity is in good agreement with the previous estimate by \citet{ols91}, 
who obtained [Fe/H]=--0.42$\pm$0.04,
while both these estimates disagree with the significant lower abundance ([Fe/H]=--0.96$\pm$0.15) 
found by H00. Unfortunately we did not re-observed the two stars
measured by  H00, hence no direct comparison can be done. However, 
the relatively large number of giants 
measured in this work and the accurate tests we perform on the abundance analysis
suggested that our result is quite solid. 
It is also worth noticing that high metallicity estimate for this intermediate-age cluster 
is in agreement with  
the recent finding \citep[see e.g.][]{cole00, smi02, cole05} that
the metallicity distribution of intermediate-age LMC
field stars 
shows a remarkable peak in the abundance distribution at [Fe/H]$\approx-0.4\pm0.2$ dex. 

Though the discussion of the overall age-metallicity relation in the LMC is
beyond the purpose of this paper, the result obtained here deserves 
a few considerations.
It is interesting to note that NGC1978 is in the
age range where different star formation (SF) models provide significantly 
different predictions in the
age-metallicity relation.  For example, the predictions of the
two models discussed by \citet{pagel} (see their Figure 4),
show significant differences for clusters in the 2-10 Gyr 
age range. The two models are also discussed by H00
 and compared with some observations (see their Figure 4a). Here we just note that
 the current age estimate 
for NGC1978 \citep[$\approx$3.5 Gyr,][]{girardi}, and our metallicity determination,
place the  cluster in a position within the age-metallicity diagram   
more consistent with a smooth
SF rather than with a bursting model.
Of course no firm conclusion can be reached on the basis of only
one cluster, however we strongly emphasize how only the  combination of accurate
metallicities and age determinations could significantly improve our knowledge
in the star formation history of the LMC. Hence   
an accurate determination of the NGC1978 age based on highly accurate
CMD is   urgent to properly locate the cluster in the 
age-metallicity diagram.
 
\begin{figure}[t!]
\epsscale{1.0}   
\plotone{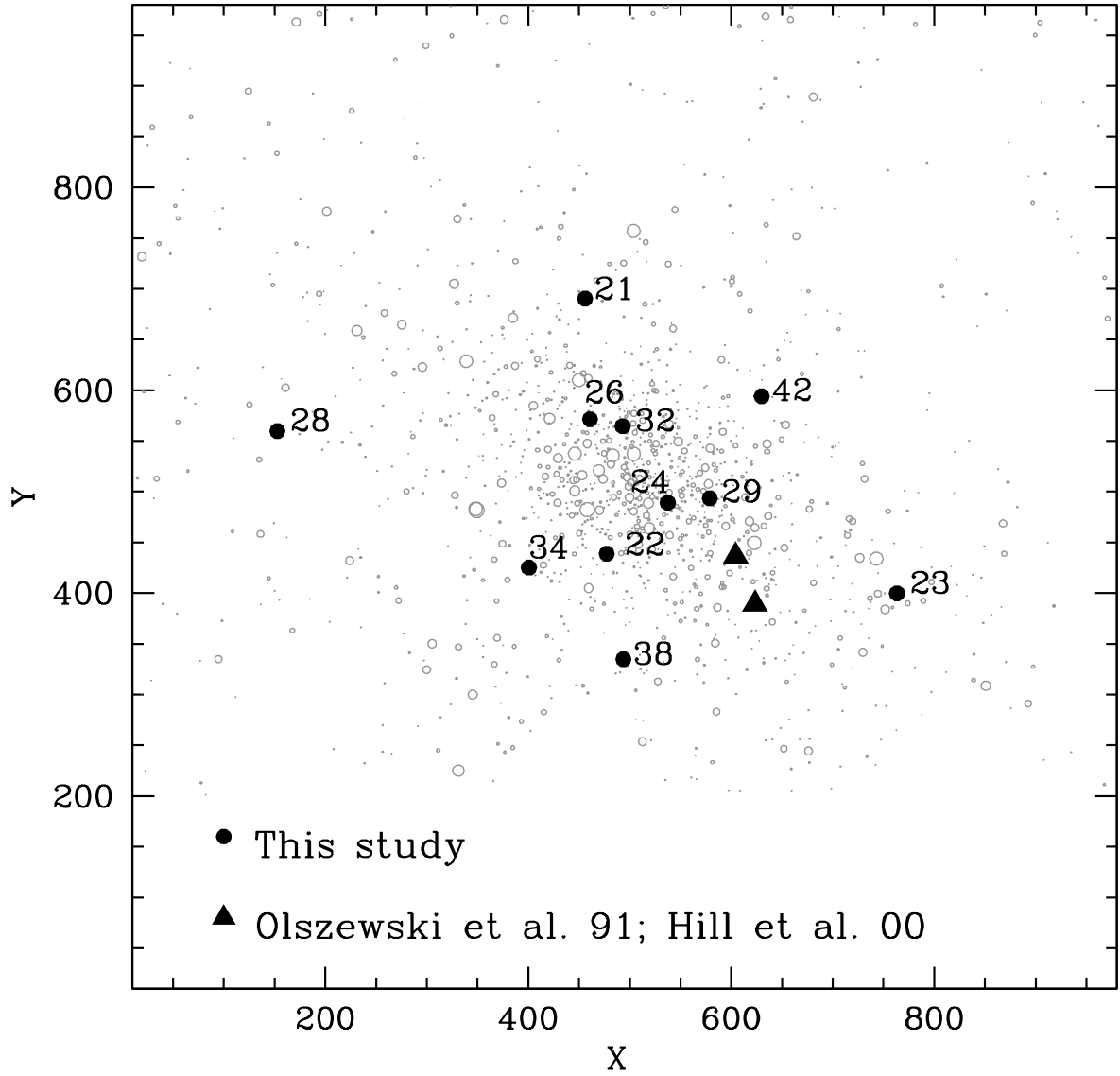}
\caption{Location of the 11 program stars (black points)  
within the cluster area.
X,Y coordinates are in pixels.
The two filled triangles mark the position of the two stars measured by \citet{ols91,hill00}.
}
\label{map}
\end{figure}

\begin{figure}[t!]
\epsscale{1.0}   
\plotone{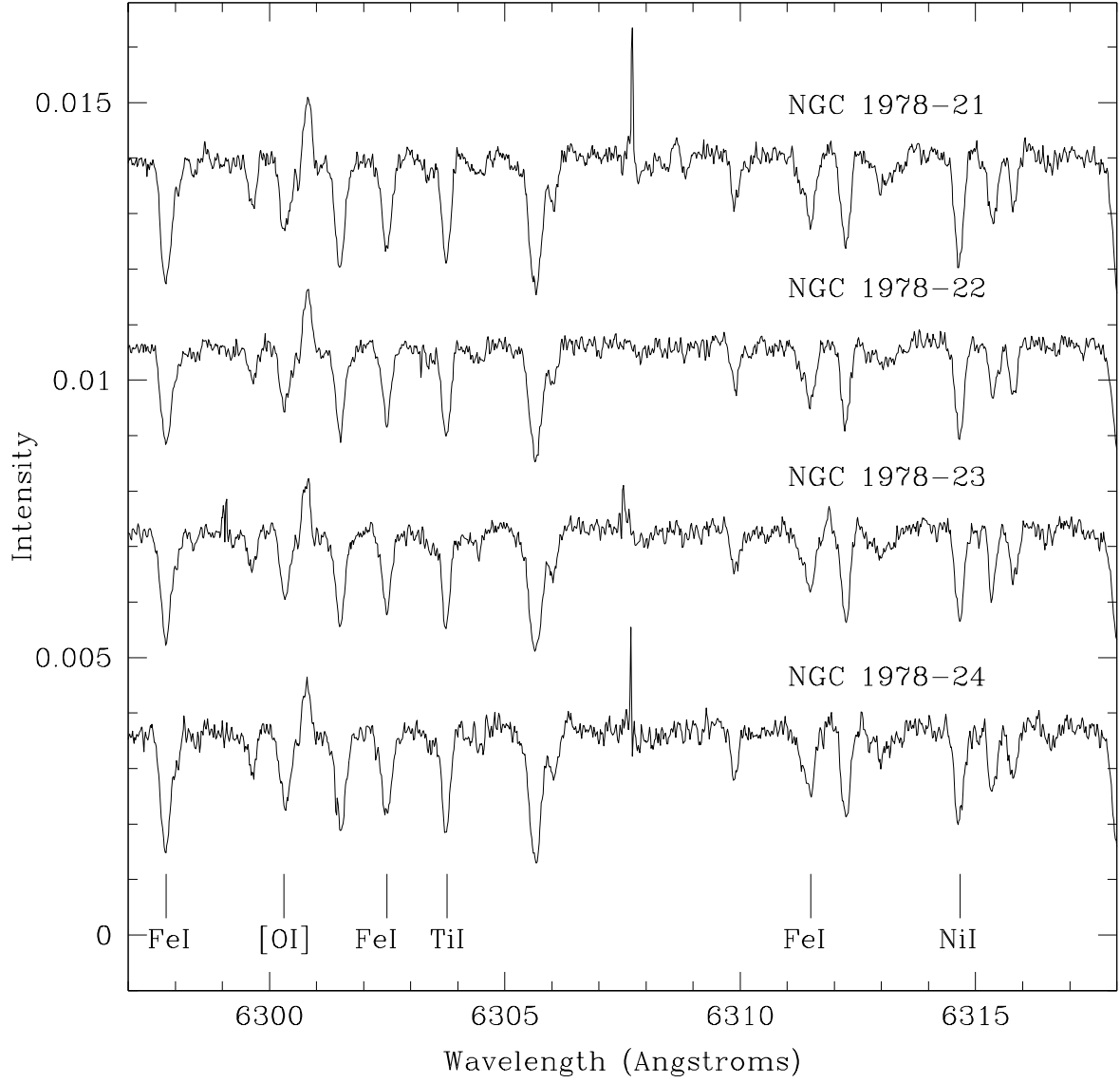}
\caption{Spectra of four program stars (typical signal-to-noise ratio is 
35-40). A few reference lines are indicated.}
\label{spec}
\end{figure}

NGC~1978 is one of the most massive stellar cluster in the LMC and it has been suspected 
to harbor a chemically inhomogeneous stellar population (see Sect.~1).
Note that both the most massive stellar systems in the halos of our Galaxy  
\citep[$\omega$ Cen, $M\sim3\cdot10^{6} M_{\odot}$,][]{MMM97} 
and M31  
\citep[G1, $M\sim7\cdot10^{6} M_{\odot}$,][]{mey01}
show evidence of 
a metallicity spread and a complex star formation history 
\citep[][]{fer04b,sol05}.  
Curiously, both these massive stellar systems show a relatively large ellipticity 
($\epsilon\approx$0.2), similar to NGC 1978. These properties
have been interpreted as possible signatures  of a merging event\footnote{Note that
 several clusters in the MC appear to be binary (or show cluster-to-cluster 
interaction).}. Hence our findings deserve a few additional comments in the
context of
the cluster formation.
The fact that our targets are well distributed within the entire
cluster area  (see Fig.~\ref{map}) and that they show an high level of homogeneity 
in their Iron abundance  
allows us to safely 
conclude that NGC 1978 does not show any signature 
of metallicity spread. 
Also, the IR CMDs presented by \citet{m06} do not confirm  
the presence of a significant spread along the RGB (contrary to the claim of \citet{alcaino}).
Of course,
our finding makes the merging hyphothesis poorly convincing
since  it would require either that  the two  sub-units had similar metallicity
or that  the two gas clouds with different metallicities efficiently mixed
at better than  $\delta[Fe/H]=0.07$ dex before star formation started.
Both these occurrences  are quite unlikely, hence we can safely conclude that
there is not signature pointing at a 
merging event in the formation history of this cluster.
Moreover, previous dynamical studies of this cluster \citep{fisher92}
already found  no evidence for merging. 
Finally, 
it is also worth noticing that ellipticity 
is a common feature of many LMC and Galactic 
clusters \citep[see e.g.][]{goodwin97} with no evidence of  a metallicity spread.
A few explanations for a large  ellipticity, other than merging, can be 
advocated, the two most likely being either cluster rapid 
rotation and/or strong tidal interactions with the parent galaxy.\\

\acknowledgements  

We warmly thank Elena Valenti and Elena Sabbi for their support during the 
preparation of the observations and data analysis, and the anonymous referee 
for his/her suggestions.
This research 
was supported by the Agenzia Spaziale Italiana (ASI) and the 
Ministero dell'Istruzione, del\-l'Uni\-versit\`a e della Ricerca.

\clearpage

\begin{deluxetable}{cccccccccccc}
\tablecolumns{12} 
\tablewidth{0pc}  
\tablecaption{Adopted atmospheric parameters and derived FeI and FeII abundances, 
for the 11 giants observed in NGC 1978.}
\tablehead{ 
\colhead{~~~Id~~~}& \colhead{$v_{rad}$}  & \colhead{$T_{eff}$}& \colhead{$log{g}$}
 &  \colhead{[A/H]} & 
\colhead{$v_t$} & \colhead{$N_{FeI}$} &\colhead{[Fe/H]I} & \colhead{rms}  
& \colhead{$N_{FeII}$} & \colhead{[Fe/H]II} & \colhead{rms}  \\
& \colhead{(km/s)} &\colhead{(K)}& \colhead{(dex)}
 &  \colhead{(dex)} & 
\colhead{(km/s)} &    
&    }
\startdata 
   1978-21 &  291.5 &   3790  & 0.64 & -0.43   & 1.54  &74&  -0.43& 0.16& ---&   --- &---  \\
   1978-22 &  290.6 &   3700  & 0.55 & -0.37   & 1.50  &78&  -0.39& 0.17&   7&  -0.27&0.19  \\
   1978-23 &  292.3 &   3630  & 0.57 & -0.24   & 1.35  &70&  -0.25& 0.21& ---&  ---&---  \\
   1978-24 &  288.7 &   3750  & 0.62 & -0.30   & 1.40  &59&  -0.30& 0.17&   1&  -0.17&---  \\
   1978-26 &  292.1 &   3820  & 0.71 & -0.43   & 1.53  &83&  -0.42& 0.15&   1&  -0.28&---  \\
   1978-28 &  290.5 &   3740  & 0.69 & -0.33   & 1.28  &85&  -0.33& 0.18&   2&  -0.17&0.01  \\ 
   1978-29 &  298.4 &   3750  & 0.71 & -0.44   & 1.58  &89&  -0.44& 0.21&   4&  -0.30&0.06  \\
   1978-32 &  291.5 &   3700  & 0.73 & -0.40   & 1.39  &84&  -0.41& 0.19&   2&  -0.30&0.18  \\
   1978-34 &  297.1 &   3900  & 0.83 & -0.32   & 1.49  &84&  -0.32& 0.20& ---&   --- &--- \\
   1978-38 &  296.3 &   3840  & 0.81 & -0.43   & 1.59  &72&  -0.44& 0.14&   2&-0.37&0.11  \\
   1978-42 &  295.5 &   3880  & 0.86 & -0.43   & 1.55  &92&  -0.43& 0.18&   2&  -0.26&0.17   \\
\enddata 
\end{deluxetable}

\clearpage 

\clearpage

\clearpage
 
\clearpage 
 
 \begin{figure}[t!]
\epsscale{1.0}   
\plotone{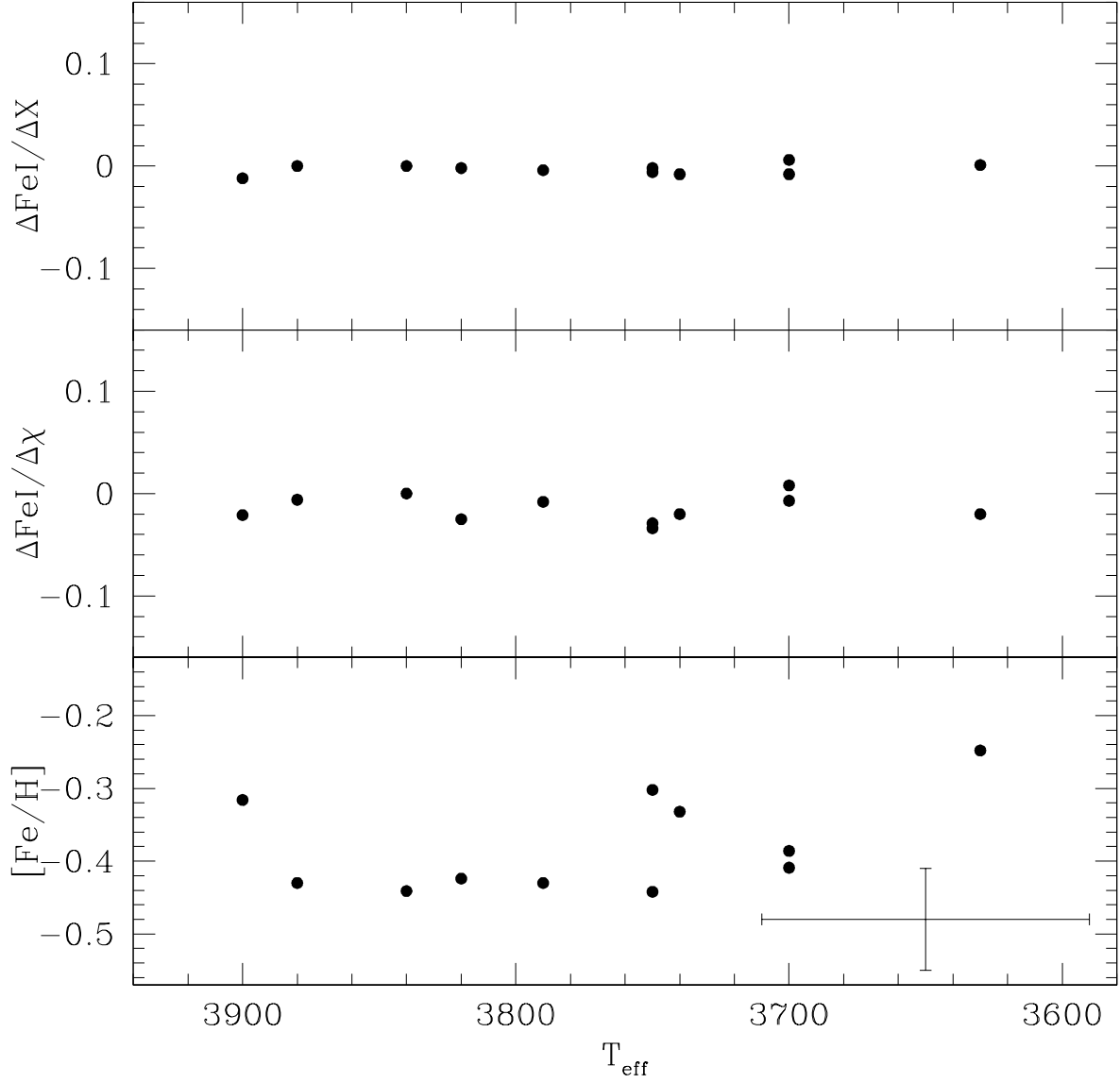}
\caption{$\Delta{FeI}/\Delta(X)$  {\it (upper panel)},  
 $\Delta{FeI}/\Delta(\chi)$ {\it (mid panel)} and  derived [Fe/H] {\it (lower panel)},
 as a function of  $T_{eff}$. The size of the typical errorbar is  shown in the lower
 panel.}
\label{temp}
\end{figure}

\end{document}